\documentclass[twocolumn,showpacs,aps,epsfig,nofootinbib]{revtex4}

\usepackage{graphicx}
\usepackage{epstopdf}
\usepackage{latexsym}
\usepackage{amssymb}

\usepackage[center]{subfigure}

\begin{document}

 \newcommand{\bq}{\begin{equation}}
 \newcommand{\eq}{\end{equation}}
 \newcommand{\bqn}{\begin{eqnarray}}
 \newcommand{\eqn}{\end{eqnarray}}
 \newcommand{\nb}{\nonumber}
 \newcommand{\lb}{\label}
\newcommand{\PRL}{Phys. Rev. Lett.}
\newcommand{\PL}{Phys. Lett.}
\newcommand{\PR}{Phys. Rev.}
\newcommand{\CQG}{Class. Quantum Grav.}

\title{Polarizing primordial gravitational waves by parity violation}

\author{ Anzhong Wang$^{1,2}$, Qiang Wu$^{1}$, Wen Zhao$^{3,1}$,  and Tao Zhu$^{1}$}
 \affiliation{${^1}$Institute for Advanced Physics $\&$ Mathematics, Zhejiang University of Technology, Hangzhou, 310032, China\\
 ${^2}$GCAP-CASPER, Physics Department, Baylor University, Waco, TX 76798-7316, USA\\
 ${^3}$Department of Astronomy, University of Science and Technology of China, Hefei, 230026, China}

\date{\today}

\begin{abstract}

We study primordial gravitational waves (PGWs)  in the Horava-Lifshitz (HL) theory of quantum gravity, in which high-order spatial
derivative operators, including the ones violating parity, generically appear  in order for the theory to be power-counting renormalizable
and ultraviolet (UV) complete. Because of both parity violation and non-adiabatic evolution of the modes due to a modified dispersion
relationship, a large polarization of PGWs becomes possible, and it could be well within the range of detection of the BB, TB and EB
power spectra of  the forthcoming  cosmic microwave background (CMB) observations.

\end{abstract}

\pacs{98.70.Vc, 98.80.Cq, 04.30.-w}

\maketitle

\section{ Introduction}

With the arrival of the precision era of cosmological measurements,   temperature and polarization
maps of  CMB  with unprecedented accuracy  will soon become available \cite{KDM}, and
shall provide a wealth of data concerning the physics of the early universe, including inflation \cite{Guth}, a dominant paradigm,
according to which primordial  density and PGWs  were created from quantum  fluctuations in the very early
universe. The former  grows to produce the observed  large-scale  structure, and meanwhile creates CMB temperature anisotropy,
which was already detected by the Cosmic Background Explorer (COBE) almost two decades  ago \cite{COBE}. PGWs,
on the other hand,  produce not only  temperature anisotropy, but also a distinguishable  signature in CMB polarization. In
particular, decomposing the polarization into two modes: one is curl-free, the E-mode, and the other is divergence-free, the
B-mode, one finds that the B mode pattern cannot be produced by density fluctuations. Thus, its detection would provide a
unique signature for the existence of PGWs \cite{seljak}. PGWs normally produce the TT, EE, BB and TE spectra of CMB, but the
spectra of TB and EB vanish when  the parity of the PGWs  is conserved.  However, if the theory is chiral,
the power spectra of right-hand and left-hand PGWs can have different amplitudes, and then induce non-vanishing TB
and EB correlation in  large scales \cite{lue}. This provides the opportunity to directly detect the chiral asymmetry of the theory
by observations  \cite{lue,saito,kamionkowski}.

With the above  motivations, the studies of PGWs  have attracted a
great deal of attention, and various aspects have been explored
\cite{KDM}. Current 7-year Wilkinson Microwave Anisotropy Probe
(WMAP) observations give the constraint on the tensor-to-scalar
ratio $r<0.36$\cite{wmap7}, and the 9-year data give the similar
result $r<0.38$ \cite{wmap9}. However, if combining with other
cosmological observations, the recent 9-year WMAP gives the
tightest constraint $r<0.13$ at $95\%$ confidence level
\cite{wmap9}, which corresponds to the amplitude of  the PGWs
$\Delta_h^2<3.03\times10^{-10}$. It should be noted that  they impose no
constraint on their chirality \cite{saito}.

In this paper, we investigate  the possibility to detect the  chirality  of PGWs through  three information channels, BB, TB and EB of CMB,
in the recently proposed  HL theory of quantum gravity, in
which parity violation happens generically \cite{horava}. Such a detection    places the theory  directly under tests, and
provides  a smoking gun in its parity violation  in the early universe. This   represents   one of few  observations/experiments that one can
construct currently as well as  in the near future, considering the quantum nature of the theory.

The HL theory  is power-counting renormalizable, because of the presence  of high-order spatial derivative operators.
The exclusion of high-order time derivative operators, on the other hand,  renders  the theory unitary, whereby it  is expected to be
UV complete. In the infrared (IR), the low-order derivative operators take over [cf. Eq.(\ref{DFE})] and presumably provides
a healthy IR limit \cite{horava}. When applying it to cosmology, various  remarkable features were found \cite{reviews}. In particular, the
higher-order spatial curvature terms can give rise to a bouncing
universe \cite{Calcagni}, may ameliorate the flatness problem \cite{KK}
and lead to caustic avoidance \cite{Mukohyama:2009tp}. The anisotropic
scaling provides a solution to the horizon problem and generation of
scale-invariant perturbations with \cite{inflation} or without inflation \cite{Mukohyama:2009gg}.
It also provides a new mechanism for generation of primordial magnetic seed field
\cite{MMS}, and a modification of the spectrum of gravitational
wave background via a peculiar scaling of radiation energy density
\cite{MNTY}. With the projectability condition, the lack of a local
Hamiltonian constraint leads to ``dark matter as an integration
constant'' \cite{Mukohyama:2009mz}. The dark sector can also have its
purely geometric origins \cite{Wanga}. A large Non-Gausianity is possible for both scalar \cite{HW2} and tensor  \cite{HWY} perturbations
even with a single slow roll scalar field, because of the presence of high order derivative terms, and so on.

Despite all of these remarkable features, the theory also faces  some challenging questions,
such as instability and strong coupling. To overcome these questions, various models have been proposed, including the ones with
an additional local U(1) symmetry \cite{HMT,zhu}, in which the problems, such as instability, ghosts, strong coupling, and different speeds in the
gravitational sector,  can be avoided. In all of those models, the
tensor  perturbations are almost the same \cite{Wang}, so without loss of the generality, we shall work with
 the model proposed in \cite{zhu}.

 The rest of the paper is organized as follows: In Sec. II we specify the model that accounts for the polarization of PGWs, while in Sec. III
 we consider the polarization of PGWs in the de Sitter background. In Sec. IV, we discuss their detectability for the Planck satellite and forthcoming
 observations. The paper is ended with Sec. V, in which we derive our main conclusions.

 It should be noted that,  although our motivations  to study the polarization of PGWs is the HL theory, our final
conclusions are applicable to any theory in which the dispersion relation of the PGWs is described by Eq.(\ref{eq14}).

In addition,   the effects of  chirality  of gravitons on  CMB
 was first studied in \cite{lue} in Einstein's theory of gravity, and lately   in \cite{soda} in the HL theory. But, our model is fundamentally
different from theirs.   In particular,   the model studied in \cite{soda}  produces a negligible polarization, and
is  not detectable within current and near future observations,
as to be shown below.  To have a sizable effect, we find that it is essential for the existence of a non-WKB region in the dispersion relation
that leads to non-adiabatic evolution of the modes, a feature that is absent
in the model of  \cite{soda}.   In addition,    chirality  of gravitons  was also considered in  \cite{Myung}, but this model
 is not power-counting renormalizable, and cannot be considered as a viable candidate of quantum gravity.

\section{The Model}

In \cite{zhu}, the parity was assumed, by setting all  the fifth- and third-order spatial derivative operators,
\bqn
\lb{Aaction}
\Delta{\cal{L}}_{V} &=&  \left(\alpha_0 K_{ij}R^{ij}   + \alpha_2{\epsilon}^{ijk}
  R_{il}\nabla_j R^{l}_{k}\right)/M_{*}^3 \nb\\
  && +  {\alpha_1} \omega_3(\Gamma)/M_{*} + ``...",
\eqn
to zero,
where $\alpha_i$'s are dimensionless coupling constants, ${\epsilon}^{ijk}$ is the total antisymmetric tensor, $K_{ij}$ and $R_{ij}$ denote, respectively,
 the extrinsic curvature and the 3-dimensional Ricci tensor  built of the 3-metric $g_{ij}$.  $\nabla_{i}$ denotes the covariant derivative with respect to
 $g_{ij}$, $\omega_3(\Gamma)$ the 3-dimensional gravitational Chern-Simos term,   $M_{*}$ the energy scale above which the high-order
derivative operators become important, and   ``..."   is the part proportional to $a_{i}$, which vanishes for tensor perturbations, where
$a_{i} \equiv N_{,i}/N$ with $N$ being the lapse function  \cite{ADM}. For detail, we refer readers to \cite{zhu}.
However, once the radiative  corrections are taken into account, these terms are expected to be present generically \cite{PS}. Therefore, in this paper
we shall add these terms into the action of \cite{zhu}, and show that it is exactly because of their presence that the PGWs will get polarized. Depending on the
strength of these  terms, the polarization of PGWs can be well within the range of detectability  of the forthcoming observations of CMB, as to be shown
explicitly below.

In the  Friedmann-Robertson-Walker   flat universe, the background is given by
$\hat{N} = a(\eta),  \;  \hat{g}_{ij} = a^2(\eta)\delta_{ij}$ and $\hat{N}^i = \hat{\varphi} = \hat{A} =0$, where $\hat{N}^i$ denotes the shift vector \cite{ADM}, and
$\hat{\varphi}$ and $ \hat{A} $ are, respectively, the Newtonian prepotential and U(1) gauge field \cite{HMT}.
 Consider the tensor perturbations,
 $\delta{g}_{ij}   =a^2h_{ij}(\eta, {\bf x})$, where $h_{ij}$ is transverse and traceless, $\partial^i h_{ij} = 0= h^{i}_{\;\; i}$.
Then, the quadratic part  of the total action  can be cast in the form,
\bqn
\lb{actionP}
&&  S_{\text{total, g}}^{(2)} = \zeta^2 \int{d\eta d^3 x}\Bigg\{\frac{a^2}{4}\left(h_{ij}'\right)^2
 - \frac{1}{4}a^2 \left(\partial_kh_{ij}\right)^2\nb\\
 && ~~ - \frac{\hat\gamma_3}{4M_*^2} \left(\partial^2h_{ij}\right)^2
  - \frac{\hat\gamma_5}{4M_*^4 a^2} \left(\partial^2\partial_kh_{ij}\right)^2\nb\\
  && ~~ -  \frac{\alpha_1 a e^{ijk}}{2M_*} \Big(\partial_{l} h^m_{i}\partial_{m}\partial_{j} h^{l}_{k}  - \partial_{l} h_{im} \partial^{l}\partial_{j}h^{m}_{k}\Big)\nb\\
  &&~~  -  \frac{\alpha_2  e^{ijk}}{4M_*^3 a}\partial^2 h_{il} \left(\partial^2 h^{l}_{k}\right)_{,j}  -    \frac{3\alpha_{0}{\cal{H}}}{8M_*^3 a} \left(\partial_kh_{ij}\right)^2 \Bigg\},
 \eqn
where $\partial^2 = \delta^{ij}\partial_{i} \partial_{j},\; {\cal{H}} = a'/a, \;  \epsilon^{ijk} \equiv e^{ijk}/\sqrt{g}$,
and $\gamma_3  = \hat{\gamma}_3 \zeta^2/M_{*}^2, \; \gamma_5  = \hat{\gamma}_5 \zeta^4/M_{*}^4$, with $\zeta^{2}= M_{pl}^2/2$
and $e^{123} = 1$.
$\gamma_{3, 5}$ are the dimensionless coupling constants of the theory \cite{zhu}, similar to $\alpha_{0, 1, 3}$.
To avoid fine-tuning,   ${\alpha}_{n}$ and  $\hat{\gamma}_{n}$  are expected to be  of  the same order.
Then,   the field equations for $h_{ij}$ read,
 \bqn
 \lb{DFE}
 h''_{ij}&+&2\mathcal{H}h'_{ij}- \alpha^2 \partial^2
 h_{ij}+ \frac{\hat\gamma_3}{a^2M_*^2}\partial^4
 h_{ij}-\frac{\hat\gamma_5}{a^4M_*^4}\partial^6 h_{ij}  \nb\\
&+ &
e_{i}^{\;\; lk}\left(\frac{2\alpha_1}{M_* a} + \frac{\alpha_2}{M_*^3 a^3}\partial^2 \right) \left(\partial^2h_{jk}\right)_{,l} = 0,
 \label{eq7}
 \eqn
where a prime denote derivative with respect to  $\eta$, $
\alpha^2 \equiv 1+ {3\alpha_0{\cal{H}}}/{(2M_*^3 a)}$.

  \section{Polarization of PGWs in the de Sitter background}

When the background is de Sitter, we have $a=-{1}/{H\eta}$, where $H$ is the Hubble constant.
To study the evolution of $h_{ij}$, we first expand it    over spatial Fourier harmonics as,
$$
 h_{ij}(\eta,{\bf x}) = \sum_{A=R,L} \int \frac{d^3 {\bf
 k}}{(2\pi)^3} \psi_k^A(\eta) e^{i{\bf k}\cdot{\bf x}}P_{ij}^A(\hat{\bf k}),
 $$
where $P_{ij}^{A}(\hat{\bf k})$ are the circular polarization tensors and
satisfy the relations: $i k_s {e}^{rsj} P_{ij}^A=k \rho^A P_{i}^{rA}$ with
$\rho^R=1$, $\rho^{L}=-1$, and ${P^*}_{j}^{iA} P_{i}^{jA'}=\delta^{AA'}$ \cite{soda}.
Substituting it into Eq. (\ref{eq7}), we obtain
 \bq
 \lb{AA}
 \phi_{k, yy}^A+ \left({\omega}_{\text{A}}^{2} -{2}{y^{-2}}\right)\phi_k^A = 0,  \label{eq13}
 \eq
 where  $\phi_k^A\equiv \sqrt{\alpha k}\; a \psi_k^{A}$, $y\equiv - \alpha k\eta$,  and
 \bqn
 {\omega}_{\rm A}^{2}
 =1+\rho^A\big(\delta_1 y +\delta_3y^3\big) + \delta_2y^2 +\delta_4y^4, ~~
  \label{eq14}
 \eqn
with $ \delta_1 \equiv - 2 ( {\alpha}_{1}/\alpha^{3})\varepsilon_{\rm HL},\;
\; \delta_2 \equiv  (\hat{\gamma}_{3}/\alpha^{4})\varepsilon_{\rm HL}^2 ,\;  \delta_3 \equiv ({\alpha}_{2}/\alpha^{5})\varepsilon_{\rm HL}^3,
\; \delta_4 \equiv  (\hat{\gamma}_{5}/\alpha^{6})\varepsilon_{\rm HL}^4$,
 where  $\varepsilon_{\rm HL} \equiv {H}/{M_*}\ll 1$. Note that  the unitarity of the theory in  the UV requires $\hat\gamma_5 > 0$,
 while a healthy IR limit requires $\alpha^2 \simeq 1$.  Thus, without loss of generality, in the following we
set $\alpha = 1$, or equivalently $\alpha_0 = 0$.
However, $\alpha_1$, $\alpha_2$ and $\hat\gamma_3$ have no such restrictions, as long as $\omega_{A}^{2} >0$ holds.
 Following \cite{martin}, we choose  the initial conditions  at $\eta = \eta_i$ as,
 \bqn
 \lb{initialC}
 \phi_k^A(\eta_i) =\sqrt{\frac{1}{2\omega_{ A}}},\; \frac{d\phi_k^A(\eta_i)}{d y}= i \sqrt{\frac{\omega_{ A}}{2}},
 \eqn
 which minimizes the energy density of the field.

Before proceeding further, let us note that the case studied in \cite{soda} corresponds to the choice,
$ \beta =  \delta_2, \; \gamma = \delta_3/(2\delta_2), \; \delta_1 = 0$ and $\delta_4 = \delta^2_{3}/(4\delta_2)$, where
$\beta$ and $\gamma$ are parameters introduced in \cite{soda}. To obtain
sizable polarization of PGWs for future observations,  \cite{soda}   showed  that
$\beta$ and $\gamma$ should be of order one, which implies  $\hat{\gamma}_{3} = {\cal{O}}\left(\varepsilon_{\rm HL}^{-2}\right),\;
{\alpha}_{2} = {\cal{O}}\left(\varepsilon_{\rm HL}^{-3}\right)$, and $\hat{\gamma}_{5} = {\cal{O}}\left(\varepsilon_{\rm HL}^{-4}\right)$. Clearly, this represents
fine-tuning.

\begin{figure}[t]
\centerline{\includegraphics[width=8cm,height=6.5cm]{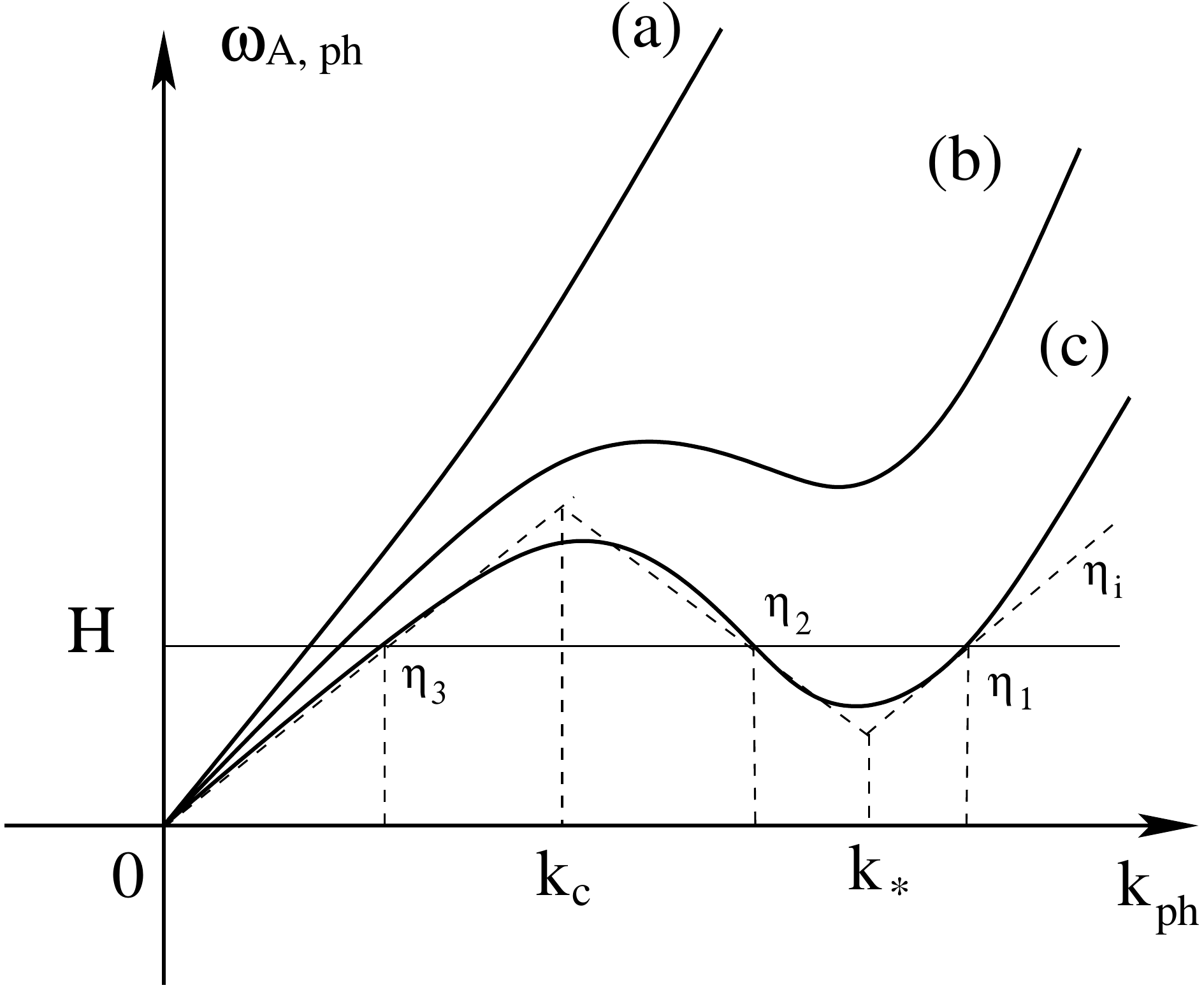}}
\caption{The evolution of $\omega_{\text{A,ph}}$ vs $k_{ph}$  for (i)  $\delta_{2} = \delta_3 = 0$ or  (ii)  $\delta_{1} = \delta_2 = 0$,
where $k_{ph} \equiv k/a$.  Region I: $\eta\in (\eta_i, \eta_1)$;
Region II: $\eta\in (\eta_1, \eta_2)$; Region III: $\eta\in (\eta_2, \eta_3)$; and Region IV: $\eta\in (\eta_3, 0)$.}
\label{fig0}
\end{figure}

To see the effects of the parity-violated terms, let us first consider  two representative cases:  (i) $\delta_2 = \delta_3 = 0$; and (ii) $\delta_1 = \delta_2 = 0$.
In each of them, we can obtain $\omega_{R, ph}$   from $\omega_{L, ph}$ by flipping the sign of $\delta_1$   or $\delta_3$, where
${\omega}_{\rm A, ph}^{2} \equiv {\alpha^2 k_{ph} ^2}{\omega_{A}^{2}}$, and $k_{ph} \equiv k/a$.
Thus, without loss of generality, we  assume that $\delta_{1, 3} \ge 0$.
Then, from Eq.(\ref{eq14}) we can see that ${\omega}_{\rm R}^{2}$ is a monotonically
 function of $y$, and  the equation, $\omega_{R, ph} =  H$, has only one real positive
root, as shown by Curve (a) in Fig. \ref{fig0}. Then, the WKB approximations are applicable in the region  $\omega_{R, ph} >  {H}$, and the mode function
$\phi^{R}_{k}  = \sqrt{\alpha k} v_k^{R}$ of Eq.(\ref{AA})  is given by
\bq
\lb{eq17b}
v^{R}_{k}  = \cases{\frac{1}{\sqrt{2\omega_{R}}}e^{-i \int_{\eta_{i}}^{\eta} {\omega_{R}(k, \eta')d\eta'}}, & $ \omega_{R, ph} >  {H}$, \cr
D_{+} a(\eta) + D_{-} a(\eta) \int_{\eta_{3}}^{\eta}{\frac{d\eta'}{a^2(\eta')}}, & $ \omega_{R, ph} <  {H}$,\cr}
\eq
where
$D_{\pm}(k)$ are uniquely determined by the boundary conditions at the horizon  crossing $\omega_R = a(\eta_3)H$, which require
$v_k^{R}$ and its first-order time derivative continuous across the boundary.

On the other hand, $\omega_{L, ph} = H$ can have one, two or three real positive roots, depending on the ratio $\delta_{1}/\delta_4$ or $\delta_{3}/\delta_4$.
If it has only one   root, as shown by Curve (b), the WKB approximations are applicable in the region  $\omega_{L, ph} > H$,
and     $v^{L}_{k}$ is given by  Eq.(\ref{eq17b}), where $\phi^{L}_{k} = \sqrt{\alpha k} v_k^{L}$.  In the case with two real roots, we have $\eta_1 = \eta_2$, and Region II
 in Fig. \ref{fig0} does not exist. As a result,  $v^{L}_{k}$  is also given by   Eq.(\ref{eq17b}).  But, when it has three
real   roots, as shown by Curve (c), the WKB approximations are not applicable in Region II, and
 the evolution  becomes non-adiabatic.  Then,  the mode function $v^{L}_{k} [= \phi^{L}_{k}/\sqrt{\alpha k}]$ of Eq.(\ref{AA})     is given by
\bq
\lb{eq17c}
v^{L}_{k} = \cases{\frac{1}{\sqrt{2\omega_{L}}}e^{-i \int_{\eta_{i}}^{\eta} {\omega_{L}(k, \eta')d\eta'}}, & Region I,\cr
 C_{+} a(\eta) + C_{-}a(\eta) \int_{\eta_{1}}^{\eta}{\frac{d\eta'}{a^2(\eta')}}, & Region II,\cr
 \frac{\alpha_k e^{-i \Theta^L_2(k, \eta)}  + \beta_k e^{i  \Theta^L_2(k, \eta)}}{\sqrt{2\omega_{L}(k, \eta)}}, & Region III,\cr
  D_{+} a(\eta) + D_{-}a(\eta) \int_{\eta_{3}}^{\eta}{\frac{d\eta'}{a^2(\eta')}}, & Region IV,\cr}
\eq
where
 $\Theta^A_n(k, \eta) \equiv \int_{\eta_{n}}^{\eta} {\omega_{A}(k, \eta')d\eta'}$. The coefficients $C_{\pm}(k), D_{\pm}(k), \alpha_k, \beta_k$ are uniquely determined
by  requiring that
$v_k^{L}$ and its first-order time derivative be continuous across   the boundaries  that separate these regions \cite{MB03}. Note that due to the non-adiabaticity of the evolution   in Region II, particles are
created,  where their occupation number $n_k$ is given by $n_k = \left|\beta_{k}\right|^2$. To have the energy density of such
particles be smaller than that of the background, one must require \cite{BM05},
 $\left|\beta_{k}\right|^2 < \left({M_{pl}}/{M_{*}}\right)^2 \varepsilon^2_{\rm HL}$.   In the case without the $U(1)$ symmetry, the study of the PPN corrections   requires $M_{*} \le 10^{15}$
 GeV \cite{BPS}.  In the case with the $U(1)$ symmetry, the spin-0 gravitons are not present, and the gravitational sector has the same degree of freedom as that in GR. So,  some softer constraint on
 the values of $M_{*}$  is expected, although such  considerations   have  been carried out so far only  for  the static spherical case \cite{LW}, in which the Eddington-Robertson-Schiff parameters were calculated,
 and found that they are consistent with observations and do not impose any constraint   on $M_{*}$.   But, it is expected that other considerations, such as frame effects, will impose some constraints
 on $M_{*}$.  With some anticipation, in this paper we simply assume $M_{*} \le M_{pl}$.
  Then, for $\varepsilon_{\rm HL} \simeq  M_{*}/M_{pl}$,  we have   $\left|\beta_{k}\right|^2 \simeq {\cal{O}}(1)$ \cite{NoteA}.

\begin{figure}[t]
\centerline{\includegraphics[width=9cm,height=10cm]{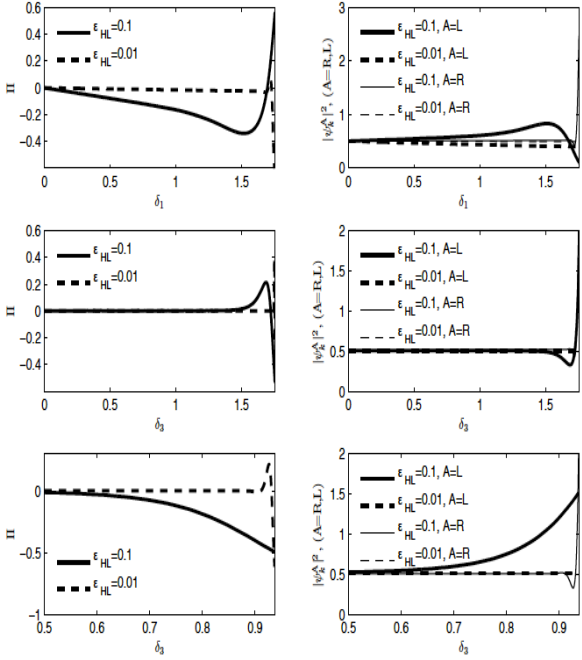}}
\caption{(a) Top panel: $\; \delta_{1} \not= 0,\; \delta_{2} = \delta_{3} = 0$.
(b) Middle panel: $\; \delta_{3} \not= 0,\; \delta_{1} = \delta_{2} = 0$. (c) Low panel:  $\; \delta_{1} = 0,\;  \delta_{2} = -  \varepsilon_{\rm HL}^{2}$.
In all the plots, we have set $
\delta_{4} = \varepsilon_{\rm HL}^{4}$.  }\label{fig3}
\end{figure}

Assuming that all the above conditions hold, we can see that  in both of  Cases (i) and (ii) there are only two distinguishable combinations,
described, respectively,  by Curves (a)+(b) and Curves (a)+(c) in Fig. \ref{fig0}. In the former,  the  power spectrum of PGWs and
the circular polarization are given by,
\bqn
\lb{eq71d}
\Delta_h^2 &\equiv&  \frac{k^3\left(|\psi_k^R|^2+|\psi_k^L|^2\right) }{(2\pi)^2}   \nb\\
 &=& \frac{H^2}{4\pi^2}\Big(1+{21 {\alpha}_1^2} \varepsilon^2_{\rm HL} +\mathcal{O}(\varepsilon_{\rm HL}^3)\Big),\\
\Pi &\equiv& \frac{|\psi_k^R|^2-|\psi_k^L|^2}{|\psi_k^R|^2+|\psi_k^L|^2}    \nb\\
&=& 3 \alpha_1 \varepsilon_{\rm HL} +\left(17 {\alpha}_1^3-3  {\alpha}_2\right)  \varepsilon^3_{\rm HL}/2 +\mathcal{O}(\varepsilon_{\rm HL}^5). \nb
\eqn
Therefore, in this case the polarization of PGWs is negligible for physically reasonable
 values of $\alpha_{1}$ and $\alpha_2$. Note that the case studied in  \cite{soda} belongs to it (with $\alpha_1 = 0$)  [cf. Fig. \ref{fig3}(b)].

For the combination of Curves (a)+(c), we find
\bqn
\lb{eq71e}
\Delta_h^2 &=&
\frac{H^2}{4\pi^2} \Big[1+\Delta_{k}^L- 3{\alpha}_1\Delta_{k}^L\varepsilon_{\rm HL} \nb\\
&& + \frac{21}{2}\left(1 + \Delta_{k}^L\right)\alpha_1^2\varepsilon_{\rm HL}^2 +\mathcal{O}(\varepsilon_{\rm HL}^3)\Big],\nb\\
\Pi &=&
-\frac{ \Delta_{k}^{L}}{1+ \Delta_{k}^{L}}
+\frac{3(1+2 \Delta_{k}^{L}){\alpha}_1}{(1+ \Delta_{k}^{L})^2}\varepsilon_{\rm HL}\nb\\
&& + \frac{9\alpha_1^2\Delta_{k}^{L}(1+2 \Delta_{k}^{L})}{(1+ \Delta_{k}^{L})^3}\varepsilon_{\rm HL}^2
+\mathcal{O}(\varepsilon_{\rm HL}^3),
\eqn
 where $ \Delta_{k}^{A}  \equiv |\beta^{A}_k|^2+\text{Re}\left(\alpha^A_k\beta^{A*}_ke^{-2 i \Theta^A_{23}}\right)$,
and $\Theta_{nm}^{A} = \Theta_{n}^{A}(k ,\eta_{m})$.
Thus,   in the present case  a large $\Pi$ becomes possible. Fig. \ref{fig3}(a)  shows such possibilities.

In addition to the above two specific  cases, when $\delta_2 \not=0$, there is another possibility in which both of $\omega^{R}_{ph}$
and $\omega^{L}_{ph}$ are given by Curve (c). Then,   we find
\bqn
\lb{eq71f}
\Delta_h^2 &=&   \frac{H^2}{4\pi^2}\Big[1+\Delta_{k}^{+} + 3{\alpha}_1\Delta_{k}^{-}\varepsilon_{\rm HL}\nb\\
&&+\frac{3}{2}\left(7{\alpha}_1^2-\hat{\gamma}_3\right)\left(1+\Delta_{k}^{+}\right)\varepsilon^2_{\rm HL}+\mathcal{O}(\varepsilon_{\rm HL}^3)\Big],\\
\Pi &=& \frac{\Delta_{k}^{-}}{1+\Delta_{k}^{+}}+\frac{3{\alpha}_1(1+2\Delta_{k}^R)(1+2\Delta_{k}^L)}{(1+\Delta_{k}^{+})^2}\varepsilon_{\rm HL}\nb\\
&&+\frac{9{\alpha}_1^2\Delta_{k}^{-}(1+2\Delta_{k}^R)(1+2\Delta_{k}^L)}{(1+\Delta_{k}^{+})^3}\varepsilon_{\rm HL}^2
+\mathcal{O}(\varepsilon_{\rm HL}^3),\nb
\eqn
where $\Delta^{\pm}_{k} \equiv \Delta_{k}^R\pm\Delta_{k}^L$.
 Again,  since   $\Delta^{A}_{k}\; (A = R, L)$  can be as large as of order one,  a large $\Pi$ now also becomes possible, as shown by Fig. \ref{fig3}(c).

 \section{Detectability of PGWs}

\begin{figure}[t]
\centerline{\includegraphics[width=8cm,height=9.5cm
]{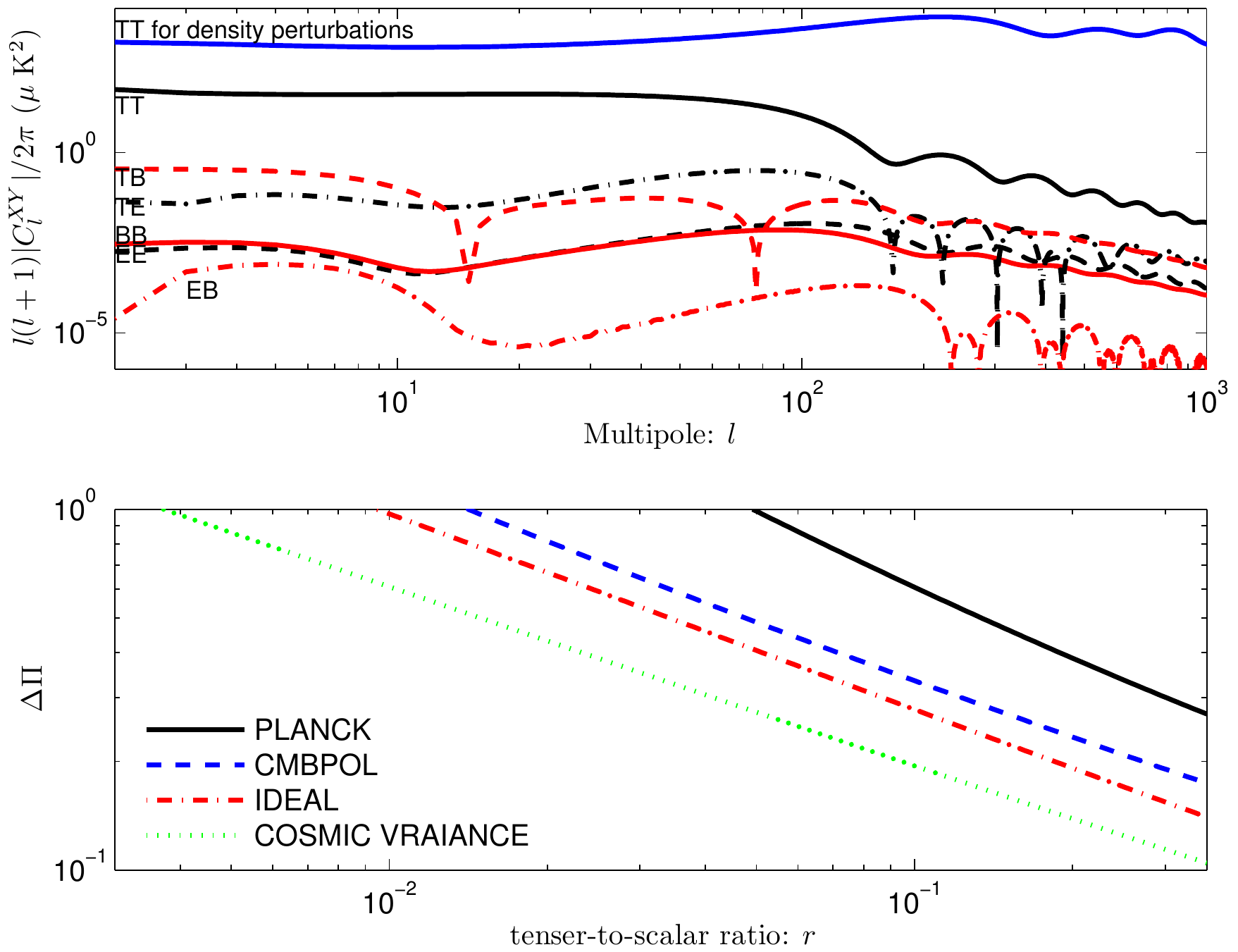}}
\caption{Top panel: The CMB power spectra generated by polarized
PGWs with $r=0.1$ and $\Pi=1$, and the TT power spectrum generated by density perturbations for comparison. Low panel: The uncertainties
$\Delta \Pi$ as a function of $r$ for potential observations.}\label{figure4}
\end{figure}

In the case of the two-point statistics, the CMB temperature and
polarization anisotropies are completely specified by six (TT, EE,
BB, TE, TB, EB) power spectra. Usually, the PGWs produce the TT,
EE, BB and TE spectra, but the spectra of TB and EB should vanish
due to the parity consideration of the PGWs. However, if the
linearized gravity is chiral as in the current case, the power
spectra of right-hand and left-hand PGWs can have different
amplitudes, and thus induce non-vanishing TB and EB correlation in
large scales \cite{lue}. This provides the opportunity to directly
detect the chiral asymmetry of gravity by observations, which has
been discussed in some detail in  \cite{lue,saito,kamionkowski}.
Different from them,   here we  consider three information
channels, BB, TB and EB, to contain the chiral PGWs by determining
both parameters $r$ and $\Pi$.  In Fig. \ref{figure4} (top panel),
we show the CMB power spectra produced by PGWs with $r=0.1$ and
$\Pi=1$, from which one can see that the PGWs are easier to be
detected in the smaller BB channel than in the larger TB one. The
main reason is,  as  stated in \cite{kamionkowski}, the
uncertainties of TB and EB channels are much larger than those of
BB channel, especially the effect of TT and EE power spectra
generated by density perturbations. So, the determination of $r$
is mainly from the BB channel, but not from TB or EB one. Note
that the background cosmological parameters are chosen as 9-year
WMAP best-fit values \cite{wmap9}, and $n_t=0$ is fixed throughout
this paper.

In order to determine the uncertainties of the parameters by the potential observations, we  use the Fisher matrix technique to avoid the Monte Carlo simulations.
The Fisher matrix is
 \bqn
 F_{ij}=\sum_{l}\sum_{XX',YY'}\frac{\partial C_{l}^{XX'}}{\partial p_i}
 {\rm C}^{-1}(D_{l}^{XX'},D_{l}^{YY'})
 \frac{\partial C_{l}^{YY'}}{\partial p_j},\nb
 \eqn
where $C_l^{XX'}$ are the CMB power spectra and $D_{l}^{XX'}$ the corresponding estimators. $p_i$ are the parameters to be determined,
which are $r$ and $\Pi$ in the present case. The covariance matrix of the estimators is given by
 $$ 
 {\rm C}(D_{l}^{XX'},D_{l}^{YY'})=\frac{\mathcal{C}_{l}^{XY}\mathcal{C}_{l}^{X'Y'}+\mathcal{C}_{l}^{XY'}\mathcal{C}_{l}^{X'Y}}{(2l+1)f_{\rm
 sky}},
$$ 
where $\mathcal{C}_{l}^{XY}=C_{l}^{XY}+N_{l}^{XY}$, and the noise
power spectra $N_{l}^{XY}$ include the instrumental noises, lensed
B-mode polarization, and the CMB power spectra generated by
density perturbations. $f_{\rm sky}$ is the sky-cut factor, which
will be taken  $f_{\rm sky}=0.65$ for Planck  \cite{planck}, $0.8$
for CMBPol \cite{cmbpol} and ideal experiments, and $1.0$ for the
cosmic variance limit. Note that for Planck and CMBPol, we have
ignored the contaminations from Galactic radiations, especially
the synchrotron and dust-dust radiations, which are expected to be
well controlled and become subdominant by the multi-band
observations \cite{foreground}. In the ideal case, only the
reduced lensed B-mode polarization is considered as the
contamination \cite{lensing}, and in the cosmic variance limit, we
assume that all the contaminations can be well removed. Throughout
our calculations, we choose $l_{\max}=2000$. Once the Fisher
matrix is calculated, the uncertainties of the parameters can be
evaluated by $\Delta p_i=\sqrt{{F^{-1}}_{ii}}$.

As mentioned in \cite{kamionkowski}, the determination of $r$ is mainly from the observation of BB information channel, but not from TB or EB channels,
even if the PGWs are completely chiral. In order to quantify the detection abilities of the experiments,
we define the signal-to-noise ratio $S/N\equiv r/\Delta r$. Similar to the discussions in the previous works \cite{zhao}, we find that if the condition  $S/N>3$ is required,
i.e. a definite detection, $r>0.03$ is needed for Planck satellite. While CMBPol mission can detect the signal  if $r>0.4\times 10^{-3}$, and the ideal experiment can detect
it if $r>0.8\times10^{-5}$ \cite{NoteB}.

The constraint on $\Pi$ is mainly from the TB and EB channels, where
the cosmic variances caused by TT, EE and BB are dominant. So,  the values of $\Delta \Pi$ are nearly independent of  $\Pi$
 in the fiducial model. In Fig. \ref{figure4} (Low panel), we present the uncertainties $\Delta \Pi$
as a function of $r$ for the four measurements, where we have set
$\Pi=1$ in the fiducial model. The results are quite close to
those presented in \cite{saito,kamionkowski}. It also shows that
if $\Delta \Pi<0.3$, the value of $r$ should be larger than $0.3$
for Planck, and $r>0.12$ for CMBPol mission, which have nearly
been excluded by the current observations \cite{wmap9}. However,
for the ideal measurement, it requires $r>0.09$. Even if we
consider the extreme cosmic variance limit, $r>0.04$ is still
required. So, we conclude that,  if $\Pi < 0.3$, the determination
of the chirality of PGWs is quite difficult, unless the
tensor-to-scalar ratio $r$ is large enough. However, if the PGWs
are fully chiral, i.e. $\Pi \sim \pm 1$, the detection becomes
much easier. We find that to get $\Delta \Pi<1$, we only need
$r>0.05$ for Planck, $r>0.014$ for CMBPol, $r>0.01$ for the ideal
experiment, and $r>0.004$ for the cosmic variance limit.

 \section{Conclusions}

In this paper,  we have studied the evolution of PGWs, described by the dispersion relation (\ref{eq14}), obtained from
the HL theory of quantum gravity \cite{zhu}. From the analytical results given by  Eqs.(\ref{eq71d}), (\ref{eq71e}) and (\ref{eq71f}),
one can see that the polarization of PGWs is  precisely due to the parity violation and non-adiabatic evolution of the mode function in
Region II of Fig. \ref{fig0}, in which particles are created, where  their occupation number is given by
$n_{k} = |\beta_{k}|^2$. Fig. \ref{fig3}, on the other hand, shows clearly  that the polarization is considerably enhanced by
 the third- and four- order spatial derivative terms of Eq.(\ref{eq14}).
The effects of the fifth-order terms were studied in \cite{soda}, and are showed explicitly in the present paper
 that their contributions to the polarization are sub-dominant,
and  are quite difficult to be detected in the near future.
The detectability of the polarization caused by other terms, on the other hand, seems  very optimistic, as shown  in  Fig. \ref{figure4} (Low panel).

It should be noted that, although the dispersion relation (\ref{eq14}) is obtained from the HL theory \cite{zhu,Wang}, our results are actually applicable to any theory where
the mode function of PGWs are described by Eqs.(\ref{eq13}) and (\ref{eq14}), including the trans-Planck physics \cite{martin}. In addition,   the  chirality  of PGWs could
also be detected by the potential lensing observations \cite{BKS}.

\section*{Acknowledgements}
This work was supported in part by DOE, DE-FG02-10ER41692 (AW), NSFC No. 11205133 (QW $\&$ TZ);
No. 11047008 (QW $\&$ TZ); No. 11105120 (TZ); No. 11173021 (WZ $\&$ AW);
No. 11075141 (WZ $\&$ AW),  and project of Knowledge Innovation Program of Chinese Academy of Science (WZ).


\begin{thebibliography}{99}
\bibitem{NoteA}  In addition, the creation of such particles also change the rolling speed of the inflaton.
To have a  sufficient long accelerating phase, one also needs to assume that
$ n_k \le  {H}/{M_{pl}}$ \cite{BM05}. This bound does not apply to the present case with two main reasons:
First,  the vacuum expectation value of $\left<\rho\right>$ and $\left<p\right>$
used  in \cite{BM05} were based on an effective  theory of a scalar field that is different from the one in the HL theory \cite{WM}.
Second, in the HL theory, as far as the horizon problem is concerned,  inflation (with slow roll conditions)
is not necessary \cite{KK}, because of the modified dispersion relation.
\bibitem{NoteB} It should be noted  that the actual detectabilities of the experiments are always worse than these evaluations,
 due to  other contaminations,
including various foregrounds, E-B leakage, systematics and so on. See for instance,  \cite{zhao}.
\end{thebibliography}
\end{document}